# The X-ray properties of the merging galaxy pair NGC 4038/9 – the Antennae

A. M. Read[1], T. J. Ponman[1] and R. D. Wolstencroft[2]
[1] *School of Physics and Space Research, University of Birmingham, Edgbaston, Birmingham B15 2TT*
[2] *Royal Observatory Edinburgh, Blackford Hill, Edinburgh EH9 3HJ*

24 May 1995

**ABSTRACT**
We report the results of an X-ray spectral imaging observation of the *Antennae* with the *ROSAT* PSPC. 55% of the soft X-ray flux from the system is resolved into discrete sources, including components identified with the galactic nuclei and large H II regions, whilst the remainder appears to be predominantly genuinely diffuse emission from gas at a temperature $\sim 4 \times 10^6$ K. The morphology of the emission is unusual, combining a halo which envelopes the galactic discs, with what appears to be a distorted, but well-collimated bipolar outflow. We derive physical parameters for the hot gas in both diffuse components, which are of some interest, given that the *Antennae* probably represents an elliptical galaxy in the making.

## 1 INTRODUCTION

The *Antennae*, NGC 4038/9, is the classic example of a pair of galaxies in gravitational interaction. Optical photographs (e.g. Arp 1966; Schweizer 1978) show spectacular tails associated with these galaxies, spanning a total angular diameter of $\sim 20'$, corresponding to 145 kpc at the distance of 25 Mpc (Tully 1988) assumed throughout this paper. The system is part of a modest concentration of galaxies called the NGC 4038 group, lying within the *Crater Cloud*, which itself lies about 7 Mpc below the Virgo Southern Extension (Tully & Fisher 1987).

The *Antennae* has been the target of a sequence of dynamical models, and as a result, its dynamical history is understood as well that of any other interacting system. Computer models such as Toomre and Toomre's (1972) simple three-body model and Barnes's (1988) N-body model in which self-gravitating halo, disc and nuclear components are taken into account, have shown that the tails can be caused by the tidal interaction, and imply that they are approximately 500 Myr old (Barnes 1988; Hut & Sussman 1987). The basic validity of these models has been confirmed by kinematical studies of both the neutral (van der Hulst 1979) and ionized (Amram et al. 1992) hydrogen gas.

As one of the closest examples of an obviously interacting system, the *Antennae* has, over the years, been the target for high resolution multiwavelength studies of the effects of tidal interaction. Optical H$\alpha$ emission knots (Rubin, Ford & D'odorico 1970) and coincident powerful radio knots (Hummel & Van der Hulst 1986; Wolstencroft 1988) are found throughout the central parts of both galaxies. Van der Hulst's (1979) 21 cm line study showed that about 70% of the H I in the system is associated with the luminous tails. Infrared images (Bushouse & Werner 1990) indicate that the nuclei of both galaxies are likely to be sites of very active star formation, and this is backed up by CO observations (Stanford et al. 1990), which show the presence of large concentrations of molecular gas, not only at the two galactic nuclei, but also at the contact region between the two discs.

The *Einstein* observation of the *Antennae*, (Fabbiano & Trinchieri 1983) was the major X-ray observation of an interacting galaxy system prior to *ROSAT*. The results were rather inconclusive, primarily due to the limited angular resolution of the *Einstein* IPC, however the emission was clearly extended, and contained a soft X-ray contribution not normally seen in spiral galaxies, which the authors interpreted as probably arising from hot gas. Harder emission and a possible hard point source in the vicinity of the contact region between the two galaxies were also seen.

The *ROSAT* X-ray telescope, with the Position Sensitive Proportional Counter (PSPC) (Pfeffermann et al. 1986) in its focal plane, offers three important improvements over the *Einstein* IPC. It has much better spatial resolution, the 90% enclosed energy radius at 1 keV being 27″ (Hasinger et al. 1992). Since, as we will see below, the source is complex, and only a few arcminutes in diameter, this improvement is critical. Secondly, the spectral resolution of the PSPC ($\Delta E/E \sim 0.4$ FWHM at 1 keV) is significantly better than previous imaging instruments, allowing characteristic source and diffuse emission temperatures to be derived. Thirdly, the internal background of the PSPC is very low ($\sim 3 \times 10^{-5}$ ct s$^{-1}$ arcmin$^{-2}$; Snowden et al. 1994), which allows low surface brightness emission to be mapped.

## 2 OBSERVATIONS, DATA ANALYSIS AND RESULTS

NGC 4038/9 was observed with the *ROSAT* PSPC on four separate occasions for a total time of over 31000 s. Each of the four datasets was reduced separately using the Starlink *ASTERIX* X-ray analysis system and they were then combined, weighting each according to its exposure time to maximise the signal/noise. Although the spectral response of the PSPC varies with time, the variation across the four *Antennae* datasets was found to have a negligible effect on the fitted models, so we were able to use the response for the longest data block for the combined data.

The data were first 'cleaned' by removing high background periods – typically 2-3% of the data. They were then binned into a spectral image (which we will also refer to as the 'data cube') covering a $0.2° \times 0.2°$ region and spanning the energy range 0.1-2.3 keV. A background model was constructed using source-free regions of an annulus situated outside the central PSPC support ring, plus knowledge of the energy dependent vignetting function, and was subtracted from the data cube.

Individual *ROSAT* pointings are subject to an r.m.s. attitude uncertainty of approximately 6″ (Hasinger et al. 1992), hence before co-adding the four component spectral images, an attempt was made to coalign them by registering each on a nearby catalogued source (the F star HD104456). Since X-ray positional errors for this source were 6-9″ (at 90% confidence), the improvement in registration was only modest, and positions from the co-added data are subject to $\approx 4″$ r.m.s. systematic errors.

From the co-added data cube, images were extracted in several different spectral bands. Fig. 1 shows contours of (0.1-1.9 keV) X-ray emission superimposed on an optical (Schmidt telescope) image, whilst Fig. 2 shows similar images in three different energy bands. Each X-ray image has been lightly smoothed with a gaussian of standard deviation 10″ to suppress noise. The contour levels in each plot increase by factors of two from $3.6 \times 10^{-4}$ ct s$^{-1}$ arcmin$^{-2}$.

The brightest X-ray emission covers the optical discs of the galaxies, with X-ray peaks centred on NCG 4038 (to the N) and NCG 4039 (to the S). These peaks are spectrally hard, whilst the remaining disc emission is more extensive in the two softer bands. A striking feature of Fig. 1 is the presence of low surface brightness X-ray features extending to the north-west and south-west, and apparently culminating in point sources (labelled P and Q in Fig. 1). The emission from these X-ray extensions is mostly very soft.

A map of the X-ray hardness is shown in Fig. 3. The grey-scale image indicates the average photon energy within each pixel containing a flux greater than $3.6 \times 10^{-4}$ ct s$^{-1}$ arcmin$^{-2}$ (i.e. corresponding to the lowest contour levels of Fig. 1). The lightest regions correspond to a mean channel energy of $\sim 0.2$ keV, the darkest, to $\sim 1.4$ keV. Whereas very hard emission is seen within the central parts of the southern disc, the central disc of NGC 4038 is rather softer (apart from a hard spot at its NE end). Much softer emission predominates over the outer parts of both discs, and also in the two X-ray extensions, though small 'hot spots' are visible.

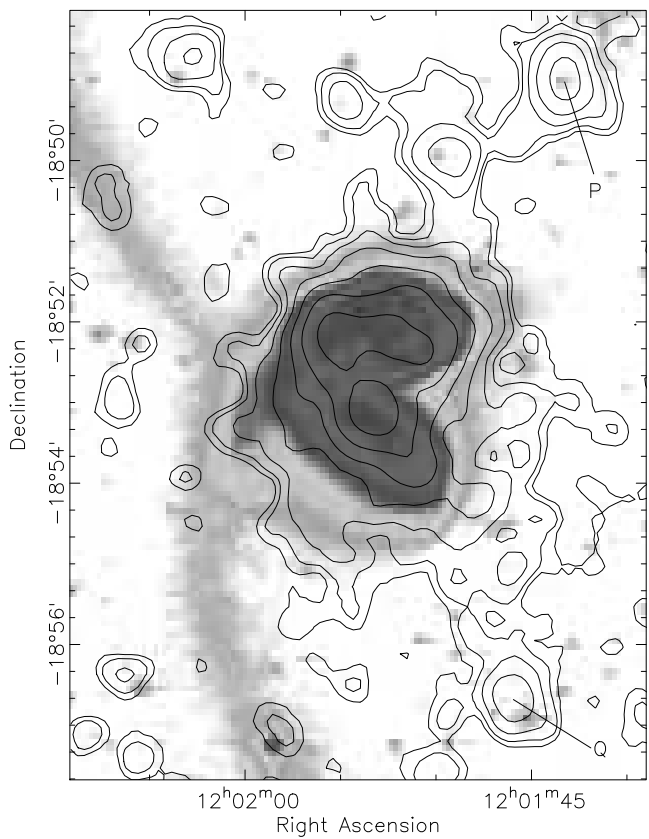

**Figure 1.** Contours of 0.1-1.9 keV X-ray emission are shown superimposed on an optical image of NGC 4038/9. The contour levels increase by factors of two from $3.6 \times 10^{-4}$ ct s$^{-1}$ arcmin$^{-2}$.

A maximum entropy reconstruction (Gull 1989; Skilling 1989) performed on a 0.1-1.9 keV image of the central regions of NGC 4038/9 and on soft (0.1-0.9 keV) and hard band (0.9-1.9 keV) images, is shown in Fig. 4. Seven discrete features (labelled A-G) can be distinguished. Best positions for these features were evaluated by locating peaks in the maximum entropy map (see Table 1).

Firstly, an integrated spectrum for the system was extracted from a circular region of radius 2.5′, and fitted with standard spectral models (hot bremsstrahlung, power-law and Raymond and Smith (1977) hot plasma). The best fit was obtained with a Raymond and Smith hot plasma model of temperature $0.69^{+0.03}_{-0.03}$ keV and metallicity $0.03^{+0.01}_{-0.01}$ times solar. The fitted column of $5.5^{+0.5}_{-0.4} \times 10^{20}$ cm$^{-2}$ is significantly greater than the column out of our own Galaxy ($3.4 \times 10^{20}$ cm$^{-2}$), indicating that either some fraction of the emission is intrinsically absorbed, or the emission is multi-component. The rather poor fit ($\chi^2 = 23.7$ with 14 degrees of freedom) suggests that a separation of the emission into its constituent components (i.e. sources and diffuse emission) is necessary.

For the purposes of further spectral analysis, the flux from point sources was removed and emission divided into two components – the bulk of the emission, which covers the two discs, and the X-ray extensions to the north and south. A spectrum of the diffuse disc emission was obtained by extracting a circular region of radius $2.5'$ containing the disc emission from the background-subtracted data cube, and removing data at the positions of the seven features A-G. The remaining data were then collapsed into a spectrum and exposure corrected. Spectra for the X-ray extensions were extracted in a similar way from approximately rectangular regions enclosing the lowest surface brightness contours shown in Fig. 1. In order to account for diffuse flux lost when removing the discrete sources, these spectra were renormalised using 'patched' images, in which the holes left at the source positions were filled by bilinear interpolation.

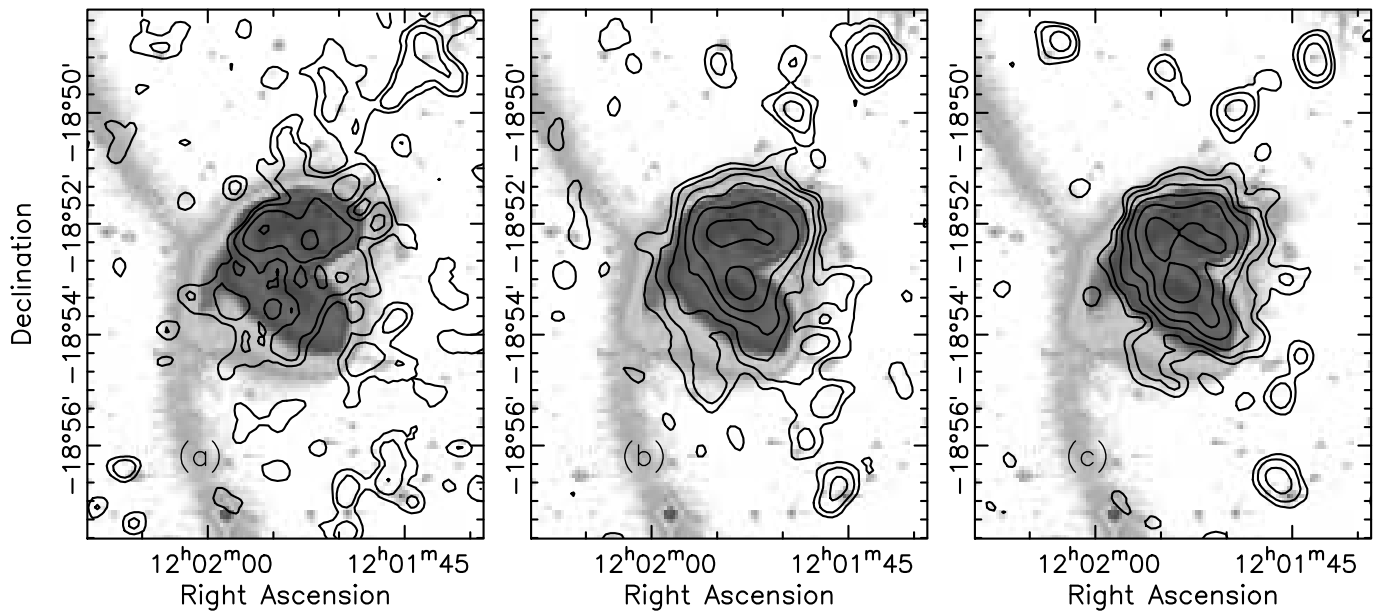

**Figure 2.** Contours of X-ray emission in three different energy bands (a: 0.1-0.4 keV, b: 0.4-0.9 keV, c: 0.9-1.9 keV) are shown superimposed on optical images of NGC 4038/9. The contour levels in each of the plots increase by factors of two from $3.6 \times 10^{-4}$ ct s$^{-1}$ arcmin$^{-2}$.

The patched image referred to above is shown in Fig. 5. The seven features A-G have been removed, as have sources detected outside the inner regions of NGC 4038/9. Separation of these features from the underlying emission proved difficult – a trade-off has to be made between removing all the source flux (implying patching to a large radius) and obtaining a reliable estimate of the diffuse flux underneath it (which may require a smaller radius if the diffuse emission is structured). Features A-G were all removed to the 95% energy enclosed radius for point sources. Sources outside the optical discs were treated similarly except in a few cases where a smaller radius was used to avoid biasing the residual diffuse flux. The final map of the diffuse emission shows that it covers the whole of both optical discs, though not the tidal arms, and extends at least $2'$ (15 kpc) beyond the optical confines of the system to the NW and SW. Within the disc emission there appear to be excesses to the west of NGC 4038 and along the line where the two galaxies make contact.

Extraction of spectra for the discrete sources A-G required some care. The 95% enclosed energy radius of the PSPC on-axis point spread function (PSF) is $\approx 0.6'$ (at $\sim 0.9$ keV), and it can be seen from Fig. 4 that circles of this size at the positions of the sources would overlap. Non-overlapping circles with radii given in Table 1 were centred on each source, and a spectrum extracted from each circle. A spectrum of the background plus diffuse disc emission was derived from the source free regions of the disc, and this was subtracted from each of the source spectra, with a normalisation determined in each case by the flux underlying each source in the patched image. Finally, the resulting spectrum for each source was scaled up to allow for the flux lost in the wings of the PSF (see Table 1). This final scaling is really energy dependent, since the PSF width varies with photon energy, being larger at high and low energies. Since we ignored this effect we will have failed to scale up the lowest and highest energy emission sufficiently, relative to the centre of the spectrum. In the worst cases, we calculate that we will have underestimated the flux by 50% at 0.2 keV, and by 20% at 2 keV. Simulations show that the low energy deficit would lead us to overestimate the column by up to 20% and the high energy deficit would cause our fitted temperature to be 10-15% too low. These worst case distortions are comparable to the statistical errors quoted in Table 2.

The spectra were fitted with standard spectral models: power law, thermal bremsstrahlung, and Raymond & Smith (1977) hot plasma models. Bremsstrahlung and hot plasma models gave results of similar quality, and the latter are generally quoted below, since they are physically better motivated. Gaussian errors are not a very good approximation for these sources, so a maximum likelihood criterion, which allows explicitly for the Poissonian nature of the statistics, was used in preference to minimum chi-squared for the spectral fitting.

The best fits to the source spectra are shown in Table 2. The comparative performance of different models can be judged on the basis of their likelihood ratios by the use of the Cash statistic (Cash 1979). In the case of the brightest sources A and D, for which the errors are fairly Gaussian, we were able to confirm that differences in the Cash statistic for different models corresponded closely to differences in chi-squared. All of the sources apart from E are better fitted by Raymond and Smith than by power law models. Where one fit is only marginally better than the other (Cash statistic values differing by $< 2$), the results for both are given, with the parameters of the poorer fit shown in brackets.

|   | RA (2000.0) | DEC (2000.0) | Source extraction radius (') | Net counts | PSF correction factor | Source idents. |
|---|---|---|---|---|---|---|
| A | 12 01 55.4 | -18 52 09 | 0.48 | 475±23 | 1.09 | H II region |
| B | 12 01 52.2 | -18 52 09 | 0.24 | 129±12 | 1.82 | nucleus (N4038) |
| C | 12 01 50.8 | -18 52 20 | 0.24 | 124±12 | 2.00 | H II region |
| D | 12 01 53.3 | -18 53 06 | 0.48 | 520±24 | 1.06 | nucleus (N4039) |
| E | 12 01 51.2 | -18 53 45 | 0.39 | 119±12 | 1.19 | H II region ? |
| F | 12 01 47.9 | -18 54 10 | 0.48 | 53±9 | 1.12 | - |
| G | 12 01 53.5 | -18 51 23 | 0.42 | 117±13 | 1.16 | - |

**Table 1.** Positions of sources A-G together with possible identifications. The radii used in the spectral extraction are given together with the net source counts within each of the circles, and the PSF correction factors used to scale up the spectra to allow for the flux lost in the PSF wings.

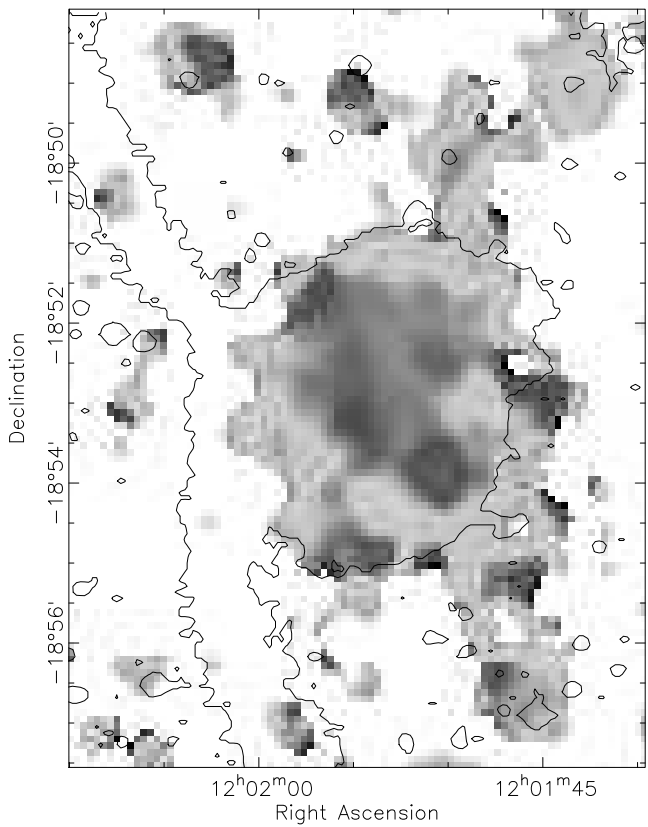

**Figure 3.** An X-ray hardness map of NGC 4038/9 superimposed on an optical outline. The average photon energy within each pixel containing a flux greater than $3.6 \times 10^{-4}$ ct s$^{-1}$ arcmin$^{-2}$ (i.e. corresponding to the lowest contour level of Fig. 1) is shown. The lightest regions correspond to a mean channel energy of $\sim 0.2$ keV, the darkest, to $\sim 1.4$ keV.

The temperatures of sources A-D are very similar and are quite well constrained (to within $\sim 0.25$ keV). Two luminosities are shown for each source; one is the source luminosity which escapes from the galaxy, the other is the *intrinsic* source luminosity, i.e. before absorption in the host galaxy. For each object, we have also calculated a 'hardness ratio', equal to the difference divided by the sum of the high- (1.0-2.5 keV) and low- (0.1-1.0 keV) energy counts. This is useful for comparison with the work of other authors. The final column of the table lists possible counterparts for some of the sources, as discussed in the next section.

The spectrum of the diffuse disc emission was also fitted with standard spectral models. The larger count rate in this case allowed the chi-squared statistic to be used. A Raymond and Smith model fitted significantly better than a simple power law spectrum ($\chi^2 = 21.6$ with 14 degrees of freedom, compared to $\chi^2 = 35.7$ with 15 degrees of freedom) and the results for this 'one component' model are shown in Table 3.

The X-ray spectra of normal spiral galaxies are rather hard, and are believed to be dominated by emission from X-ray binaries (Fabbiano 1989). One expects that a similar contribution, due to unresolved hard sources in the bulges and discs of the two galaxies, should be present in the *Antennae*. To investigate this, the diffuse spectrum was also fitted with a two component model, a Raymond and Smith hot plasma model (representing the truly diffuse emission) plus a hot (10 keV) bremsstrahlung model (representing the contribution from unresolved X-ray binaries). The diffuse component column was constrained to be no less than the column out of our own Galaxy ($3.4 \times 10^{20}$ cm$^{-2}$). The best fit model parameters are shown in Table 3, and this model is overlaid on the diffuse spectrum in Fig. 6, whilst Fig. 7 shows confidence contours for temperature against metallicity of the Raymond and Smith component.

In Table 3, the total 0.1-2.0 keV luminosity of the diffuse disc emission is shown for both emission leaving the galaxy (escaping), and that emitted within the host galaxy (intrinsic). Only the hot bremsstrahlung component of the two-component fit is affected by the host column (its fitted column is significantly greater than the column out of our own Galaxy).

The spectra extracted for each of the X-ray extensions were also fitted with standard spectral models, with the column frozen at the Galactic value. Both northern and southern features were best fitted with hot plasma models, with very similar temperatures to the diffuse disc emission (see Table 4).

|   | Hardness ratio | Spectral fits | | | Photon index | Log $L_X$ (erg s$^{-1}$) (0.1-2.0 keV) | |
|---|---|---|---|---|---|---|---|
|   |   | Column $10^{20}$ cm$^{-2}$ | Temp. keV | Z (solar) |   | escaping | intrinsic |
| A | $-0.21\pm0.05$ | $8.1^{+1.4}_{-1.1}$ | $0.72^{+0.13}_{-0.10}$ | $0.00^{+0.01}_{-0.00}$ | -   | 40.28 | 40.57 |
| B | $-0.29\pm0.09$ | $5.1^{+1.6}_{-1.3}$ | $0.79^{+0.13}_{-0.13}$ | $0.06^{+0.09}_{-0.04}$ | -   | 39.99 | 40.14 |
| C | $-0.12\pm0.10$ | $17.4^{+14.4}_{-7.4}$ | $0.68^{+0.26}_{-0.26}$ | $0.02^{+0.06}_{-0.02}$ | (4.3) | 39.91 | 40.40 |
| D | $-0.24\pm0.05$ | $13.7^{+5.8}_{-3.4}$ | $0.70^{+0.06}_{-0.10}$ | $0.10^{+0.07}_{-0.05}$ | -   | 40.16 | 40.51 |
| E | $-0.02\pm0.09$ | $10.5^{+15.8}_{-4.0}$ | (1.44) | (0.00) | $2.2^{+1.1}_{-0.5}$ | 39.70 | 40.11 |
| F | $-0.72\pm0.21$ | $1.9^{+5.6}_{-1.9}$ | $0.45^{+0.71}_{-0.17}$ | $0.02^{+0.64}_{-0.02}$ | (3.9) | 38.98 | 38.98 |
| G | $-0.39\pm0.11$ | $11.5^{+20.4}_{-4.5}$ | $0.38^{+0.17}_{-0.15}$ | $0.00^{+0.02}_{-0.00}$ | (5.8) | 39.42 | 39.98 |

**Table 2.** Best fits to the spectra of discrete sources within NGC 4038/9, with one-sigma errors for one interesting parameter. Source spectra A, B, C, D, F and G are best fitted by Raymond and Smith hot plasma models. Source spectrum E is best fitted by a power-law model. Both escaping and intrinsic source luminosities are shown. The 'hardness ratio', is the difference divided by the sum of the high (1.0-2.5 keV) and low (0.1-1.0 keV) energy counts. Luminosities are based on an assumed source distance of 25 Mpc.

|   | Spectral parameters | One-component model | Two-component model |
|---|---|---|---|
|   | Net counts (scaling factor) | $521\pm32$ (1.78) | |
| Raymond | Column ($10^{20}$ cm$^{-2}$) | $3.30^{+0.87}_{-0.77}$ | $3.40^{+0.58}_{-0.00}$ |
| and Smith | Emission measure ($10^{60}$ cm$^{-3}$) | $9190^{+3600}_{-2600}$ | $9020^{+2500}_{-1900}$ |
| plasma | Temperature (keV) | $0.49^{+0.07}_{-0.06}$ | $0.36^{+0.09}_{-0.05}$ |
| (RS) | Metallicity (Solar) | $0.04^{+0.02}_{-0.02}$ | $0.07^{+0.05}_{-0.03}$ |
| Hot | Column ($10^{20}$ cm$^{-2}$) | - | $87.5^{+215}_{-87.5}$ |
| Brems. | Emission measure ($10^{60}$ cm$^{-3}$) | - | $2200^{+4500}_{-1200}$ |
| (Br) | Temperature (keV) | - | 10.0 (Frozen) |
| $\chi^2$ (No. degrees of freedom) | | 21.56 (14) | 17.80 (12) |
| $\log L_X$ (escaping) | | 40.63 | RS comp: 40.61<br>Br Comp: 39.51<br>Total : 40.64 |
| $\log L_X$ (intrinsic) | | 40.63 | RS comp: 40.61<br>Br Comp: 40.20<br>Total : 40.75 |

**Table 3.** Best fit parameters and luminosities resulting from fitting the diffuse disc emission spectrum. Also given is the number of counts extracted for the diffuse disc spectrum and the factor by which this is scaled to allow for the diffuse flux lost when removing the discrete sources (see text). Errors are one-sigma for one interesting parameter.

Finally, we have investigated the spectral properties of the outlying sources P and Q. Power law models gave the best fits to both spectra, though thermal spectra are not ruled out. The best fit parameters are shown in Table 5.

## 3 DISCUSSION

### 3.1 The nuclei

Millimetre wave CO observations show large quantities of molecular gas in the nuclei of both galaxies (Stanford et al. 1990): $8\times10^8 M_\odot$ in NGC 4038, and $2\times10^8 M_\odot$ in NGC 4039. Optical and IR images indicate that both are sites of active star formation (Bushouse & Werner 1990), and star formation rates of $\approx 2 M_\odot$ yr$^{-1}$ have been derived for both nuclei from corrected H$\alpha$ fluxes (Stanford et al. 1990). They are also unusually bright in the radio – their extension and flat indices indicating a starburst origin for the emission, with inferred supernova rates of 0.14 yr$^{-1}$ and 0.07 yr$^{-1}$ respectively (Hummmel & Van der Hulst 1986).

The X-ray feature B appears to be associated with the above-mentioned powerful radio, CO and H$\alpha$ features observed at the centre of NGC 4038 (e.g. source 7 of Wolstencroft 1988). However, it does show a significant offset – the centre of feature B lies 10″ west of the (Stanford et al. 1990) CO feature and 13″ west of the (Hummel & van der Hulst 1986) radio feature. Similarly, feature D appears to be associated with the powerful multi-wavelength sources at the centre of the southern galaxy (e.g. source 1 of Wolstencroft 1988). Again offsets between the different features are observed, though they are somewhat smaller (feature D lies 9″ north of the (Stanford et al. 1990) CO feature but only 5″ north of the (Hummel & van der Hulst 1986) radio feature).

|  | Net counts | Spectral fits | | | Log $L_X$ (erg s$^{-1}$) (0.1-2.0keV) |
|---|---|---|---|---|---|
|  |  | Column $10^{20}$ cm$^{-2}$ | Temp. keV | Z (solar) |  |
| Northern extension | 80±18 | 3.4 (Frozen) | $0.44^{+0.51}_{-0.21}$ | $0.00^{+0.08}_{-0.00}$ | 39.30 |
| Southern extension | 83±17 | 3.4 (Frozen) | $0.43^{+0.23}_{-0.12}$ | $0.05^{+0.18}_{-0.04}$ | 39.32 |

**Table 4.** Best fits to the northern and southern extension spectra with one-sigma errors for one interesting parameter. In both cases, Raymond and Smith hot plasma models gave the best fit. Luminosities are based on an assumed distance of 25 Mpc.

|  | RA (2000.0) | DEC (2000.0) | 90% error radius (″) | Net counts | Spectral fits | | | Log $L_X$ (erg s$^{-1}$) (0.1-2.0keV) (intrinsic) |
|---|---|---|---|---|---|---|---|---|
|  |  |  |  |  | Column $10^{20}$ cm$^{-2}$ | Photon index | Temp. keV |  |
| P | 12 01 43.3 | -18 48 56 | 4.7 | 113±12 | $8.3^{+3.4}_{-3.0}$ | $4.1^{+0.9}_{-0.9}$ | (0.37) | 40.92 |
| Q | 12 01 46.0 | -18 56 37 | 4.8 | 68±10 | $1.8^{+3.1}_{-1.8}$ | $1.7^{+1.0}_{-0.9}$ | (4.36) | 39.38 |

**Table 5.** Best fit parameters for power-law fits to the spectra of sources P (to the north) and Q (to the south). Bremsstrahlung models gave marginally worse fits and the resulting temperatures are shown in brackets. Luminosities are based on an assumed source distance of 25 Mpc. Errors are one-sigma for one interesting parameter.

The X-ray spectra of both 'nuclear' components are best fitted by absorbed, low temperature plasma models. The level of intrinsic absorption seen, however, is low compared to the optical extinction inferred for the nuclear H$\alpha$ emission of $\sim 2 - 3$ magnitudes, equivalent to a column $N_H \sim 5 \times 10^{21}$ cm$^{-2}$. This, together with the $\sim 1$ kpc offsets in X-ray emission compared to CO and radio emission, suggests that we are not seeing the nuclei themselves in X-rays but instead are observing hot gas near the nuclear regions. Such extranuclear hot gas might be expected given the large star-formation rates and extended flat index radio features (indicative of a great deal of supernova activity) observed in the nuclei. The fact that the southern nucleus has the higher soft X-ray luminosity, whilst the northern one contains more molecular gas and has a higher star formation rate and radio flux, also supports the idea that we are not seeing the central starbursts directly.

The soft spectra of these features (hardness ratio $\sim$ -0.2 to -0.3) are comparable to those of superbubbles seen in the LMC (Williams & Chu 1995), though their luminosities are two orders of magnitude higher.

### 3.2 Emission from H$_{II}$ regions

The star-forming knots seen in H$\alpha$ (Rubin et al. 1970) are predominantly found in NGC 4038 and correspond to massive ($\sim 10^7 M_\odot$) H$_{II}$ regions. The total star-formation rate over all the knots can be estimated by combining the calculated nuclear star-formation rates of Stanford et al. (1990) with the total H$\alpha$ luminosity one expects to observe based on the thermal radio emission (Hummel & van der Hulst 1986). This global star-formation rate comes to $\sim 20 M_\odot$ yr$^{-1}$, an order of magnitude more than the total in our own Galaxy. Corresponding knots are seen in the radio (see Wolstencroft 1988) and contain about 35% of the total 1.5 Ghz emission (Hummel & van der Hulst 1986). Some of these radio knots are actually brighter than those at the galactic cores.

Both X-ray features A and C seem coincident with optical and radio counterparts which are most readily interpreted as large H$_{II}$ regions. The centre of source A, the brightest of the X-ray features, lies within the largest of the H$\alpha$ knots (source E of Rubin et al. 1970), and $\sim$8″ northeast of the centre of radio knot 5 (Hummel & van der Hulst 1986). Source C lies very close (within 2-3″) to radio knot 10 (Hummel & Van der Hulst 1986) and H$\alpha$ knot T (Rubin et al. 1970). The radio sources associated with A and C have the flattest radio indices of all the observed radio knots, implying the highest ratio of thermal to non-thermal emission (Hummel & van der Hulst 1986). This suggests strongly that the X-ray emission from A and C arises from collections of supernova remnants. This idea is further supported by the fact that a comparison of the radial profiles of features A and C with the Point Spread Function (PSF) of the *ROSAT* PSPC indicates that both features appear marginally extended on a scale of between 1.5 and 4.5 kpc. However, given the presence of underlying diffuse emission, higher resolution X-ray observations (which should be available soon from the *ROSAT* High Resolution Imager) are required to confirm this.

Again, as with features B and D, the hardness ratios of these features can be compared with ratios of known features in the LMC and in M101 (Williams & Chu 1995). As with features B and D, the ratios are low (though not as low), and are comparable with features such as the old, extended supernova remnant, N44, in the LMC. The luminosities of features A and C are higher, by a factor of 3-30 than the brightest features seen in M101, and, if we have truly resolved features A and C, they are also larger (the H$_{II}$ regions within M101 have diameters up to 1 kpc (Kennicutt 1984)).

Feature A may well be associated with the near-IR bar-like feature extending from the northern nucleus to the south-east to radio knot 5 (close to X-ray source A) (Wright

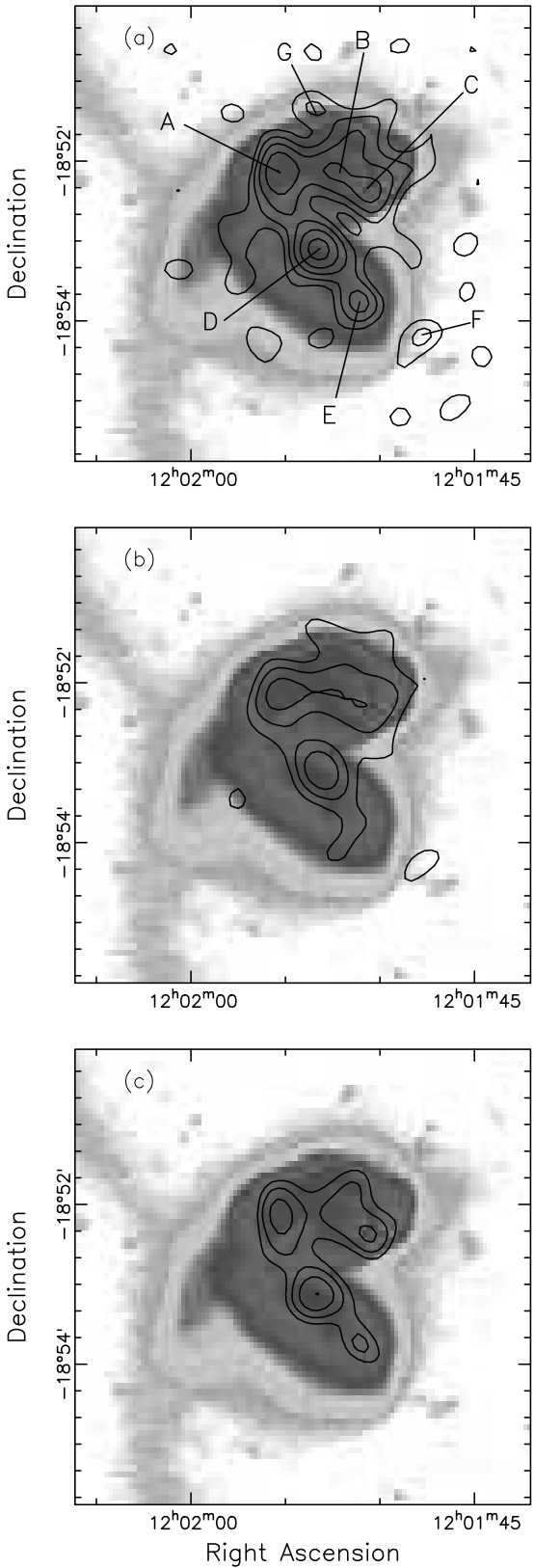

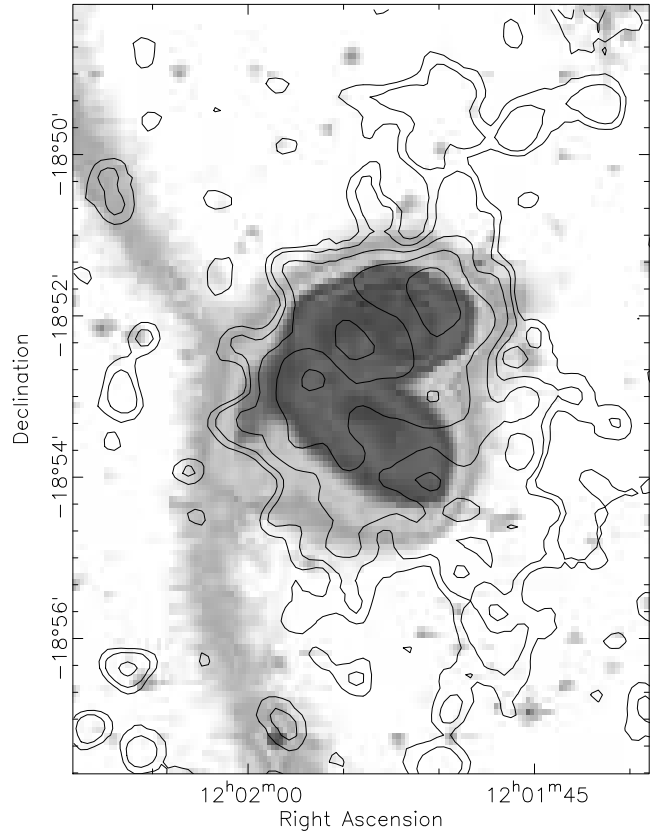

**Figure 5.** Residual (0.1-1.9 keV) diffuse emission from NGC 4038/9 after removal of sources, and linear interpolation over the resulting 'holes'. Contour levels increase by factors of two from $3.6 \times 10^{-4}$ ct s$^{-1}$ arcmin$^{-2}$.

**Figure 4.** Maximum entropy reconstructions of NGC 4038/9 in three different energy bands (a: 0.1-1.9 keV, b: 0.1-0.9 keV, c: 0.9-1.9 keV) superimposed on optical images. The contour levels increase by factors of two.

& Mclean 1987). It is suggested that the non-axisymmetric gravitational potential associated with this bar might be responsible for driving gaseous material towards the nucleus. It is now well established that substantial starburst emission can occur at the *ends* of bars as well as at the nucleus (e.g. Knapen & Beckman 1994).

### 3.3 Other discrete X-ray features

Sources E, and F, lying south-west of the southern nucleus, and source G, lying approximately due north of the northern nucleus, are all apparently, and intrinsically, less luminous than sources A-D discussed above. Source E is the only one of these with a possible counterpart at other wavelengths – it may be associated with a weak optical knot (source AA) noted by Rubin et al. (1970).

Sources E and G lie within the optical confines of the galaxies, and the fits to the spectra of both indicate absorbing columns $\sim 10^{21}$ cm$^{-2}$), well above the column out of our Galaxy. The most likely candidates for foreground Galactic sources at these flux levels, late type stars, can therefore be ruled out, since their X-ray spectra do not show intrinsic absorption. It is also very unlikely than E and G could be background objects (e.g. quasars) shining through the entire H<small>I</small> column of the *Antennae*. Using the quasar $\log N - \log S$ function of Hasinger et al. (1993), the probability of even one background quasar appearing within the optical confines of the system at flux levels as high as those observed, through an absorbing column consistent with the fitted columns of sources E and G is only 0.02.

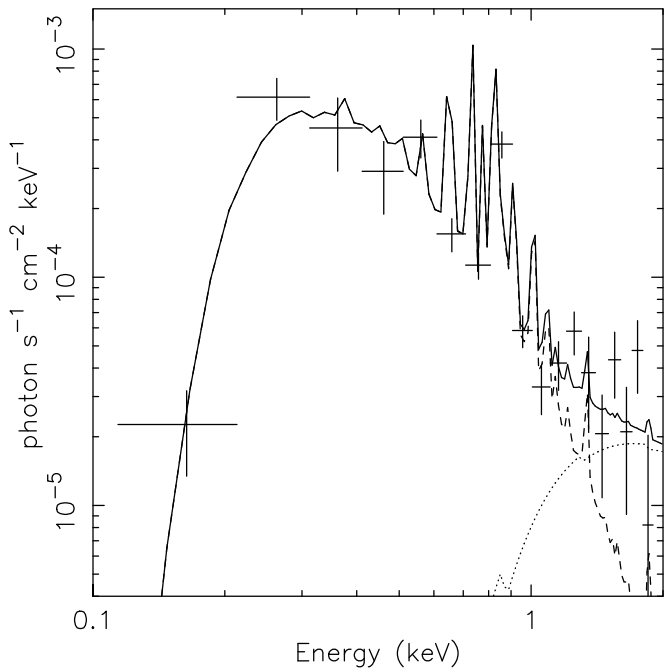

**Figure 6.** Diffuse emission spectrum from NGC 4038/9 with best fit two-component model (solid line). The contributions of the two components are shown as dashed (Raymond and Smith) and dotted (hard bremsstrahlung component) lines.

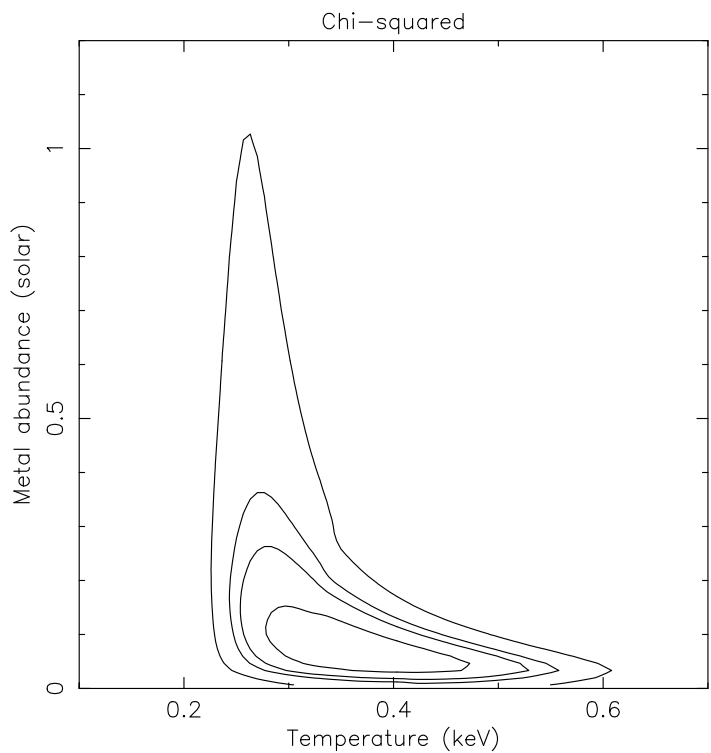

**Figure 7.** 99%, 95%, 90% and 68% confidence contours in the temperature-metallicity plane for the Raymond and Smith component of the two-component fit to the diffuse disc spectrum (see Table 3). All other parameters (except the hot bremsstrahlung temperature, which is frozen at 10 keV) are left free to optimise.

It seems likely then, that E and G are sources within the *Antennae*. Source E is actually the hardest of all the point sources, and stands out clearly in Fig. 3. It is not as hard as known X-ray binaries within the LMC, e.g. LMC X1 and R136 (Williams & Chu 1995), but is harder than the superbubble features discussed earlier, and may therefore be a collection of hot supernova remnants with possibly some contribution from X-ray binaries. Approximately 50-100 sources similar to LMC X-1 would be needed to explain the high X-ray luminosity of source E. Source G is likely, given its low fitted temperature, to be a superbubble or a collection of supernova remnants, though its X-ray luminosity is some two orders of magnitude greater than superbubbles in the LMC (Williams and Chu 1995).

Source F lies outside the optical confines of the system and has no optical counterpart (the nearest optical source with a magnitude $< 20^m$ is almost $0.5'$ away). The X-ray source is weak, and its spectrum is poorly constrained, but clearly is very soft. Bright 'supersoft' sources, which may have luminosities in excess of $10^{38}$ erg s$^{-1}$ have been detected by *ROSAT* in a number of nearby galaxies (e.g. Kahabka, Pietsch & Hasinger 1994), but in the present case the fact that the source lies just outside the visible disc of NGC 4039 makes such an identification unlikely. The soft spectrum is consistent with a foreground white dwarf, but this would be expected to be brighter than $\sim 17^m$ in the optical. Hence, assuming that the feature *is* a point source, it seems most likely (given that it lies beyond the absorbing effects of the galactic discs) to be an unusually soft, optically faint background quasar.

### 3.4 Residual disc emission

The residual emission covering the optical discs, after removal of features A-G, constitutes $\sim$40% of the total flux and appears to envelope the whole system apart from the tidal tails. Some structure in this emission can be seen in Fig. 5. The excess emission on the western side of NGC 4038 follows closely the arc-like structure of knots seen in both the radio (Hummel & van der Hulst 1986) and H$\alpha$ (Rubin et al. 1970), and peaks close to H$\alpha$ knot G (Rubin et al. 1970) and radio knot 13 (Hummel & van der Hulst 1986). In addition, there appears to be enhanced emission stretching east-west along the interface between the two galactic discs and peaking at the eastern end.

The best one-component fit to the residual disc spectrum is a low temperature (0.49 keV) Raymond and Smith plasma, absorbed by a column approximately equal to the Galactic hydrogen column ($3.4 \times 10^{20}$ cm$^{-2}$). Adding an additional 10 keV bremsstrahlung component to represent unresolved hard sources improves the fit somewhat (Table 3). However, as can be seen from Fig. 6, this hard component only contributes significantly towards the top of the PSPC energy range, and an F-test shows that the improvement in fit quality is significant at less than 90% confidence. Hence although such a hard contribution is to be expected, it is not clear that we have detected it. The error region for the parameters of the hard component allowed by our data are correspondingly large (Table 3) – a higher luminosity is allowed provided that the intrinsic absorption associated with the component is increased, such that it continues to have little impact in the *ROSAT* energy range. For the best fit model, the hard component contributes ∼8% of the 'diffuse' disc flux, and has an intrinsic 0.1-2.0 keV luminosity of $1.6 \times 10^{40}$ erg s$^{-1}$, and a bolometric luminosity ∼ 3.5 times larger. This compares, for example, to a total luminosity in discrete X-ray sources of ∼ $2.6 \times 10^{39}$ erg s$^{-1}$ in our own Galaxy (Watson 1990).

The luminosity of the soft diffuse disc component (which is affected very little by uncertainties over the hard component) is $4.1 \times 10^{40}$ erg s$^{-1}$ in the 0.1-2.0 keV band.

### 3.5 The collision interface

Direct evidence for interaction of the two galaxies in seen in the *Antennae* in the form of a large molecular gas complex (seen in CO emission) and a peak in the diffuse radio flux (Stanford et al. 1990; Hummel & van der Hulst 1986) at the interface between the discs. The high magnetic field (≈ 40 $\mu$G) implied by the radio observations probably arises from compression of the ISM. The enhancement in the residual X-ray emission between the two discs seen in Fig. 5 has a maximum at the eastern end, close to the CO and diffuse radio peaks, and is therefore also probably associated with the collision. This is not the same as the much more luminous hard source tentatively identified with the contact region by Fabbiano & Trinchieri (1983) on the basis of an observation with the *Einstein* IPC. The higher spatial resolution of *ROSAT* shows that this hard emission actually originates predominantly from the powerful, absorbed source D, at the nucleus of NGC 4039.

The hardness map, Fig. 3, shows that the emission from the interface region is harder than the bulk of the diffuse disc emission. An X-ray spectrum was extracted from an area of 0.6 arcmin$^2$ centred on the 'contact peak', and fitted with a hot plasma model, freezing the column at the value obtained from fitting the whole of the diffuse disc emission ($3.3 \times 10^{20}$ cm$^{-2}$). This resulted in a temperature of $0.94^{+0.07}_{-0.08}$ keV, significantly hotter than the remaining diffuse emission. If this is interpreted as collisionally heated gas, the temperature implies a shock speed of ∼ 700 km s$^{-1}$. Fig. 3d of Barnes (1988) indicates that the two galaxies should have a radial velocity difference of 300-400 km s$^{-1}$. Assuming a similar tangential component (not unreasonable, given the modelled orientation and history of the system), this translates into a collision velocity of 450-550 km s$^{-1}$, in good agreement with inferred collision velocities for 'typical' interacting systems (∼500 km s$^{-1}$; Harwit & Fuller 1988). Alternatively, this 'hot spot' might be due to enhanced star formation taking place in the dense molecular gas near the contact region.

### 3.6 The X-ray extensions

The extensions of the diffuse emission to the NW and SW of the system which can be clearly seen in Figs.1 and 5, look rather like X-ray analogues to the tidal arms. However they apparently have no counterparts at other wavelengths. This implies that they cannot have been separated from the main body of the system by gravitational forces, and cannot arise from unresolved discrete sources.

The idea that we are seeing streamers of hot gas is supported by the spectral properties of the X-ray emission (see Table 4). This hot gas has presumably been ejected from the body of the system by gas pressure forces. Two possible sources of energy for this are potentially available – injection from stars and supernovae, and the kinetic energy of the galaxy-galaxy encounter.

Starburst activity such as is seen in the *Antennae* is known in many cases to trigger galactic winds: massive galactic-scale outflows of hot gas driven by supernovae and, to a lesser extent, stellar winds (Heckman, Lehnert & Armus 1993). The extended soft X-ray emission seen above the discs of many starburst galaxies both with *Einstein* (e.g. Watson, Stanger & Griffiths 1984; Fabbiano 1988) and *ROSAT* (Heckman 1993; Read 1994) are thought to be due to ambient interstellar gas clouds in the galactic disc and halo, which have been shock-heated by a hot, tenuous wind.

The temperatures and luminosities inferred for the X-ray extensions in the *Antennae* are similar to what is seen in the winds of the nearby starburst galaxies NGC 253 and NGC 3628 (Read 1994). However, the soft X-ray emission is visible to a larger radius (approximately double) in the *Antennae*. This may reflect the fact that the hot wind in the *Antennae* has encountered, and shock heated, cool gas which had already been ejected from the galactic discs as a result of the violent encounter. The presence of such ejecta is supported by the observation of a considerable amount of H I beyond the optical confines of the galaxies (van der Hulst 1979), and tidally stripped material apart from the main tidal tails is also seen in dynamical models of the system (Hut & Sussman 1987; Barnes 1988). A hot galactic wind, travelling at a velocity of ∼ 2000 − 3000 km s$^{-1}$, can easily reach and shock high altitude clouds within the timescale (∼ $10^8$ yr) since pericentre indicated by modelling (Barnes 1988).

The morphology of the X-ray extensions in the *Antennae* differs from that of the classic bipolar galactic winds which emanate perpendicularly to the galactic disc in systems such as M82 and NGC 253. Although the extension consists of two branches, these are not oppositely directed, but at an angle of ∼ 120°. However the *Antennae* is a very rapidly evolving system. Over the past $2 − 3 \times 10^8$ years it is believed that NGC 4039 has swung round from the north and passed in front of NGC 4038 to reach its current position to its south. Substantial slewing of the inner galactic discs of both galaxies has also occurred. It is not too surprising then, if the wind departs from the standard bipolar form. We cannot even be certain, given the fact that there are *two* active star-forming nuclei, that we are not seeing two separate winds.

The other possibility for powering a hot gas outflow is the direct thermalisation of some of the kinetic energy of the two colliding galaxies. Harwit et al. (1987) have invoked direct collisions between molecular clouds to explain the very high infrared luminosities of some merging galaxies. On a larger scale, hydrodynamical models of merging galaxy clusters have shown (Evrard 1990) that gas is shock heated along the interface between the intracluster gas associated with each cluster, and that this hot, overpressured gas can then squirt out perpendicular to the collision axis. In order for such a phenomenon to explain the rather linear feature seen in the *Antennae*, we would have to be viewing the disc of squeezed gas approximately edge-on. This conflicts with the picture which emerges from the dynamical modelling of the system (Barnes 1988), which implies that the angle between the viewing direction and the collision axis is quite small ($\sim 30°$).

### 3.7 A pair of satellite sources

One striking aspect of the X-ray extensions seen in Fig. 1 is the fact that both appear to culminate in bright sources. These sources, which we designate P (to the NW) and Q (to the SW) are symmetrically disposed at a distance of $\sim 20$ kpc from the optical boundary of the galaxies. Both have radial profiles consistent with the point spread function of the PSPC and have no catalogued counterparts. Distortion of the X-ray contours due to these sources can be seen in the *Einstein* image (see the figure in Fabbiano, Kim & Trinchieri 1992), though they have never been commented upon.

We have obtained an optical image from the photographic co-addition of 5 sky-limited IIIaJ Schmidt plates of the *Antennae* field to search for optical counterparts to the two sources. A stellar image with B$\approx$19.8 lies within the 90% confidence error circle of P, and can be seen in Fig. 1, however nothing is visible within the error circle of Q down to B$\approx$24.

Source P has an unusual spectrum, lacking in both high and low energy flux, which is fitted by a steep spectrum with an absorbing column which exceeds that out of our Galaxy (see Table 5). Source Q has a flatter spectrum and smaller column. We derived fluxes for the two sources at each of the four epochs at which our data were collected. The resulting light curves are shown in Fig. 8. The errors plotted are purely statistical, but since the sources are not significantly contaminated by diffuse emission we expect systematic effects to be negligible. Fitting the light curves to constant flux values results in a reduced $\chi^2$ of $> 4$ for both P and Q. Both sources appear to have varied in brightness by a factor of a few on a timescale of $\sim 1$ yr. This confirms that they must be point-like.

A rather similar situation has recently been noted in NGC 4258 by Pietsch et al. (1994), who detect two bright pointlike X-ray sources, located symmetrically at a distance of 9 kpc from the weak Seyfert nucleus, and coaligned with the tips of a pair of 'anomalous arms' within the galactic disc. A mechanism for ejecting massive ($\sim 10^8 M_\odot$) objects from galactic nuclei has been suggested by Saslaw, Valtonen & Aarseth (1974). If P and Q represented ejected objects, and if the ejection process could also give rise to a trail of hot gas, this would explain the curious juxtaposition of the sources and the X-ray extensions.

Unfortunately, it is difficult to make this interesting idea work, given the substantial width ($\sim 7$ kpc) of the extensions. A very high velocity projectile might generate a broad feature after adiabatic expansion from a narrow hot wake, but no temperature gradient is seen along the extensions (see Fig. 3). Alternatively, ejected objects might create a channel out of the system through which hot gas generated by starburst activity could escape as a wind. However, the fact that the point sources appear at the *ends* of the X-ray extensions has no natural explanation within such a model.

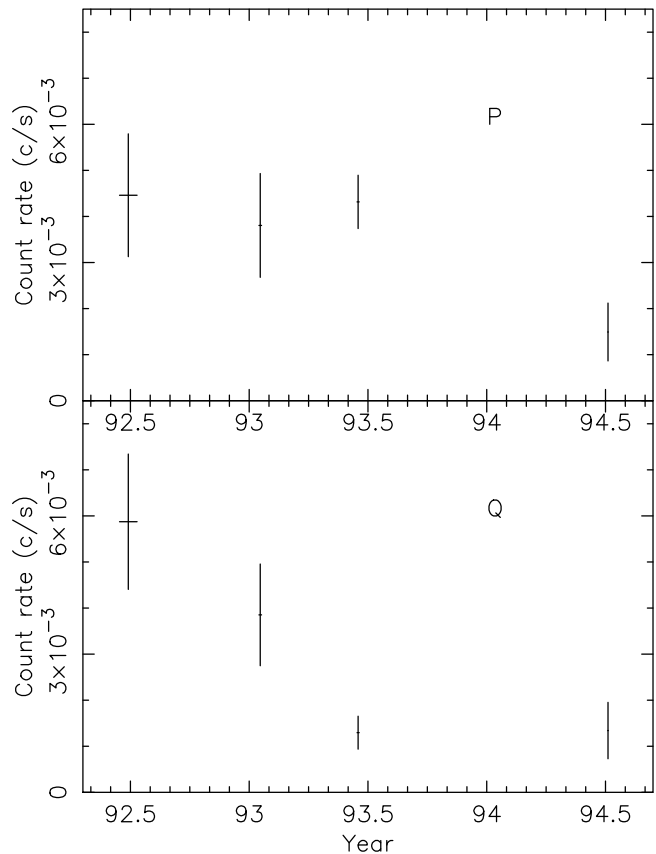

**Figure 8.** Variation in count rate from sources P and Q over the four separate observations. Errors are $1\sigma$.

Despite their suggestive symmetry and positioning relative to the X-ray 'arms', P and Q might be entirely unrelated background or foreground sources. In this case what could they be? The most common sources in ROSAT fields at these flux levels are quasars. The mean spectrum of quasars in the *ROSAT* band can be represented by a power law with a photon index of $2.2 \pm 0.2$ (Branduardi-Raymont et al. 1994; Roche et al. 1995), though some individual spectra are known to be shallower ($\alpha \sim 1.6 - 1.7$; Bühler et al. 1995). This is consistent with the spectral properties of source Q (its best fit column is low, but is consistent with the galactic value within 90% confidence), though it would be an unusually optically faint example (L.R.Jones, private communication). The X-ray spectrum of P is steeper than that of quasars. Its fitted temperature (0.37 keV) and possible optical counterpart are consistent with a late-type star, but the inferred column is too large. Optical spectroscopy of both sources is required to make further progress.

### 3.8 A tidal dwarf galaxy

Mirabel, Dottori & Lutz (1992), have discovered a dwarf galaxy in the process of forming near the tip of the southern tidal tail of the *Antennae*. The nascent galaxy consists of a chain of nebulae ionized by very young ($< 2 - 6$ Myr) massive stars, embedded in an envelope of H I gas and low surface brightness optical emission.

This region lies outside the field shown in Fig. 1, but well within the PSPC field. We detected no significant X-ray emission from this region, and derived a $2\sigma$ upper limit to the count rate within a circle of radius $0.6'$, (the 95% enclosed energy radius of the PSPC point spread function for a 0.5 keV source) at the position of the massive ($\sim 10^9 M_\odot$) H I complex at the tip of the tidal tail (van der Hulst 1979). Assuming a 5 keV bremsstrahlung model and Galactic absorption, this upper limit corresponds to a 0.1-2.0 keV luminosity $L_X < 4.3 \times 10^{38}$ erg s$^{-1}$ (the equivalent 1 keV bremsstrahlung upper limit is $4.0 \times 10^{38}$ erg s$^{-1}$). Including an additional intrinsic column, due to the local H I would raise these limits only slightly, since the column through the H I complex is only $1-2 \times 10^{20}$ cm$^{-2}$.

Dwarf irregular galaxies have a wide range of X-ray luminosities, but the fact that we detect no X-ray emission in this case, despite the active star formation, is not surprising if all the stars have been formed within the last 6 Myr (Mirabel et al. 1992). Even the most massive of these stars will not have had time to evolve off the main sequence into the X-ray bright supernova remnant or high-mass X-ray binary phase. From Mirabel et al. (1992), one can estimate a visual luminosity $L_V$ for the dwarf galaxy of $\approx 3.8 \times 10^{41}$ erg s$^{-1}$. This gives to rise to an upper limit to $L_X/L_V$ of $1.1 \times 10^{-3}$. O stars have a maximum $L_X/L_{bol}$ ratio of only $10^{-5.44}$ (Chlebowski, Harnden & Sciortino 1989), and so would not be detectable in our X-ray data.

### 3.9 Physical properties and history of the hot gas

Mean physical properties for the hot gas can be inferred from our spectral fits if we make some assumptions about the geometry of the emission. Considering first the diffuse emission covering the discs of the two galaxies, this occupies a roughly circular region on the sky, and the simplest assumption to make is that it in fact constitutes a spherical bubble of the radius observed ($\sim$15 kpc). This is the largest volume over which the gas is likely to be distributed. A radial profile of the X-ray surface brightness, after removal of the discrete sources (i.e. profiling the image in Fig. 5), shown in Fig. 9 casts some doubt on the validity of the spherical bubble model. The surface brightness falls rapidly outside the region covered by the optical discs, suggesting that the diffuse emission may be bound to the individual discs in the form of hot coronae. Assuming an isothermal gas at the observed temperature ($4 \times 10^6$ K)) and adopting an analytical Plummer-Kuzmin form (Binney & Tremaine 1987) for the potential of each disc, one derives a scale height of approximately 4.8 kpc for the gas above each disc. This agrees quite well with the scale on which the surface brightness drops off outside the disc region in Fig. 9. Taking this as the depth of the gas layer along the line of sight, and assuming a simple cylindrical geometry gives the minimum volume for the emitting gas.

Using the volumes derived for the bubble and coronal models, the fitted emission measure $\eta n_e^2 V$ (where $\eta$ is the 'filling factor' – the fraction of the total volume $V$ which is occupied by the emitting gas) can be used to infer the mean electron density, $n_e$, and hence the total mass $M_{gas}$, thermal energy $E_{th}$ and cooling time $t_{cool}$ of the gas, and also the mass cooling rate $\dot{M}_{cool}$ and adiabatic expansion timescale $t_{exp}$.

Similar estimates were made (assuming uniform density and cylindrical geometry in the plane of the sky) for the X-ray extensions to the north and south. All the resulting parameters are listed in Table 6.

In the case of the X-ray extensions, the morphology strongly suggests that there is some sort of collimated flow. If $\eta \sim 1$ for the X-ray emitting gas, then they are presumably streamers of gas at $T \sim 4 \times 10^6$ K flowing away from the system. Unless some additional force is confining them, these streamers will expand at their sound speed ($\sim 300$ km s$^{-1}$) and dissipate on a timescale $\sim 3 \times 10^7$ yr unless they are continually replenished. Alternatively we might be seeing soft X-ray emission from clouds with $\eta \ll 1$ which are being shock heated by a fast collimated wind. Theoretical modelling of starburst winds tends to support this option (Suchkov et al. 1994), since it suggests that the soft X-ray luminosity of such clouds should considerably exceed that of the hot wind fluid itself.

Turning to the gas which covers the discs, if this is even approximately in hydrostatic equilibrium, then its filling factor is unlikely to be much less than unity, since it would then require hotter low density gas to fill the rest of the volume, and this could not be bound to the galaxies (which have typical escape temperatures of $\sim 10^7$ K). The alternative is again that we are seeing gas clouds which are shock heated by an escaping hot wind, invisible in our waveband, in which event $\eta \ll 1$. Since the diffuse disc emission appears to envelope the whole system, this would require either that the wind is directed quite precisely towards us, or that it is essentially spherical. The former case seems unlikely given the far from face-on aspect of the galactic discs, whilst for a spherical wind it is hard to understand the abrupt cutoff in emission at a radius of $\sim 15$ kpc, except in the direction of the two extensions. This cannot be due to a lack of material in the vicinity available for shocking by the wind, since for example there is plenty in the nearby tidal tails. Hence one would have to suppose that the hot spherical wind does not extend beyond the X-ray bright region. Since its velocity is expected to be $\sim 2000 - 3000$ km s$^{-1}$, this implies that it must have been initiated within the past $10^7$ yr, which seems unlikely since the interaction timescale is an order of magnitude longer.

Assuming, as seems probable, that the diffuse disc gas is bound to the system and has $\eta \sim 1$, its total mass, $\sim 1 - 1.5 \times 10^9 M_\odot$, is substantial. For comparison, the main disc region contains $\sim 2 \times 10^9 M_\odot$ of molecular gas (Stanford et al. 1990) and $\sim 1 \times 10^9 M_\odot$ of H I (van der Hulst 1979). Assuming that this gaseous halo has been generated over the past $1 - 2 \times 10^8$ yr, the gas loss rate is $\sim 8 M_\odot$ yr$^{-1}$ whilst cooling is currently returning the gas to the central regions at only a fraction of this rate. The energy requirement involved in generating this halo of hot gas is essentially equal to its thermal energy content (the gravitational potential energy adds a further $\sim$10%), which corresponds to the energy released in a few million supernovae. These figures may be compared with estimates for the total star formation rate in the system of $20 M_\odot$ yr$^{-1}$, and for the supernova rate of $\sim 1.4$ yr$^{-1}$, about 15% of which takes place in the nuclei (the supernova rate can be calculated using the rates given in van den Bergh & Tammann (1991) and scaling these with the far-infrared luminosity as in Devereux & Eales (1989)). Hence, mass and energy requirements of producing such a halo do not present a problem.

| | Total diffuse disc emission | | Northern elongation | Southern elongation |
|---|---|---|---|---|
| | (bubble) | (coronal) | | |
| $\log L_X$ (erg s$^{-1}$) | 40.61 | | 39.30 | 39.33 |
| $T_X$ (keV) | 0.36 | | 0.44 | 0.43 |
| $n_e$ (cm$^{-3}$) | $0.0049 \times 1/\sqrt{\eta}$ | $0.0077 \times 1/\sqrt{\eta}$ | $0.0066 \times 1/\sqrt{\eta}$ | $0.0042 \times 1/\sqrt{\eta}$ |
| $M_{\rm gas}$ ($M_\odot$) | $1.5 \times 10^9 \sqrt{\eta}$ | $9.7 \times 10^8 \sqrt{\eta}$ | $7.1 \times 10^7 \sqrt{\eta}$ | $9.0 \times 10^7 \sqrt{\eta}$ |
| $E_{\rm th}$ (erg) | $3.1 \times 10^{57} \sqrt{\eta}$ | $2.0 \times 10^{57} \sqrt{\eta}$ | $1.7 \times 10^{56} \sqrt{\eta}$ | $2.2 \times 10^{56} \sqrt{\eta}$ |
| $t_{\rm cool}$ (Myr) | $2410 \sqrt{\eta}$ | $1530 \sqrt{\eta}$ | $2700 \sqrt{\eta}$ | $3260 \sqrt{\eta}$ |
| $\dot{M}_{\rm cool}$ ($M_\odot$ yr$^{-1}$) | 0.62 | 0.63 | 0.03 | 0.03 |
| $t_{\rm exp}$ (Myr) | 59 | 16 | 29 | 30 |

**Table 6.** Values of physical parameters for the diffuse disc gas (using both hemispherical bubble and coronal models), and for the northern and southern extensions (see text). $\eta$ is the filling factor of the gas.

The cooling time of the halo is over $10^9$ yr, which suggests that old merger remnants might retain an X-ray bright halo for some time after the optical merger (believed in a merger such as the *Antennae*'s to result in an elliptical galaxy (Toomre 1977)) is complete. Such halos are, of course, commonly observed in elliptical galaxies (Fabbiano 1989), but are attributed to material ejected during the early star forming history of early type galaxies. The halo we see in the *Antennae* differs significantly from those seen in elliptical galaxies. Fig. 9 shows that the diffuse X-ray flux falls sharply outside the disc region, with surface brightness $S(r) \propto r^{-4.5}$, and is contained within a radius $r \sim 25$ kpc, apart from the X-ray extensions. In contrast, elliptical halos have X-ray surface brightness profiles which are much flatter (typically $S(r) \propto r^{-2}$), extend over radii $\sim 50-100$ kpc, and contain $\sim 10^9 - 10^{10}$ $M_\odot$ of gas (Fabbiano 1989; Forman et al. 1993; Trinchieri et al. 1994; Rangarajan et al. 1995). The temperature of the hot halos in ellipticals is $T \approx 0.7-1.2$ keV (Matsushita et al. 1994; Rangarajan et al. 1995), considerably higher than the $T \approx 0.4$ keV seen in the *Antennae*.

These differences do not preclude the possibility that the corona seen in the *Antennae* might evolve into a more massive, hotter and extended halo if more energy and mass is injected into it as the merger proceeds, over the next few $\times 10^8$ yr. As is well known from the study of galaxy clusters (e.g. Gorenstein et al. 1978), an isothermal halo of hot gas in hydrostatic equilibrium in a potential well has a density profile, and hence an X-ray surface brightness profile, determined by the ratio of its temperature to the virial temperature in the potential. The low temperature of the corona at present therefore leads directly to a steep profile, and the injection of extra specific energy, raising the temperature to values similar to those seen in ellipticals, would result in a much flatter profile (doubling the temperature would cause the surface brightness index to drop by almost a factor of two).

It is also interesting to compare our results with the properties of the X-ray faint, early-type galaxy NGC 4365 (Fabbiano, Kim & Trinchieri 1994; Pellegrini & Fabbiano 1994), which is believed, on account of its counter-rotating core, to be the product of a recent merger. A very soft ($T \approx 0.2$ keV) component is observed in the X-ray spectrum of the system, accounting for about half of the (0.1-2.0 keV) flux. There is tentative evidence that the X-ray surface brightness extends beyond the optical profile, to a radius of $\sim 20$ kpc, and Pellegrini & Fabbiano (1994) suggest that the observed soft spectral component arises from a combination of stellar emission from the galaxy, and a gaseous halo with $T \approx 0.6$ keV, and $L_X \sim 2 \times 10^{40}$ erg s$^{-1}$. If this interpretation is correct, then the temperature and extent of this halo are comparable to what we see in the *Antennae*. Its luminosity is somewhat lower, but NGC 4365 is also an optically fainter system.

The mass and thermal energy content of the X-ray extensions is considerably smaller than that of the halo gas even if $\eta \sim 1$ for this gas. However this is probably gas which is being lost entirely from the system. If we are seeing the wind itself (i.e. $\eta = 1$) then the mass loss rate can be calculated from the inferred gas density, assuming a wind velocity. This gives a mass loss rate, including both extensions, of $\dot{M} \sim 15 \sqrt{\eta} v_{1000} M_\odot$ yr$^{-1}$, where $v_{1000}$ is the velocity in units of 1000 km s$^{-1}$, which could significantly deplete the interstellar gas content of the galaxies if it continued for $\sim 10^8$ yr. Similarly, the rate at which kinetic energy is lost in the wind is $L_{\rm wind} \sim 5 \times 10^{42} \sqrt{\eta} v_{1000}^3$ erg s$^{-1}$ (both these results are maxima in the case where the extensions lie perpendicular to our line of sight). Hence a significant fraction of the total supernova luminosity of $\sim 4 \times 10^{43}$ erg s$^{-1}$ may be carried off in the wind. If the hot wind itself is invisible to *ROSAT* then the above results still apply for the mass and energy lost in the clouds (where now we will have $\eta \ll 1$), provided that they do escape from the system. However, there is additional mass and energy loss associated with the hotter gas.

As is shown in Table 3, the metallicity of the hot disc gas inferred from hot plasma fits is very low, and the results of fits to the X-ray extensions given in section 3.6 show the same phenomenon. This result seems very surprising given the high supernova rate inferred in the *Antennae* and the consequent injection of metal-rich gas expected. It is necessary to be cautious in interpreting metallicity results from *ROSAT* spectra, since the low spectral resolution makes it impossible to distinguish individual lines, rather one is primarily fitting the effects of a blend of lines (mostly iron L lines) at $E \sim 1$ keV. Where the emission actually consists of a mixture of gas phases spanning a range of temperatures, a single temperature fit can result in an underestimate of the metallicity. Similar low abundances have been seen in diffuse emission from a number of normal spiral galaxies (Read 1994), and in the case of elliptical galaxies, it has been reliably established by much higher spectral resolution studies with the ASCA observatory, that the abundance of the gas in hot X-ray emitting halos is $\lesssim 0.5$ solar in many cases (Loewenstein et al. 1994; Matsushita et al. 1994). If the low abundance inferred for the gas in the *Antennae* is correct, then the X-ray emitting gas cannot represent the supernova ejecta, but must be derived from relatively uncontaminated gas from the galactic halos, or almost primordial material in the galaxies' environment.

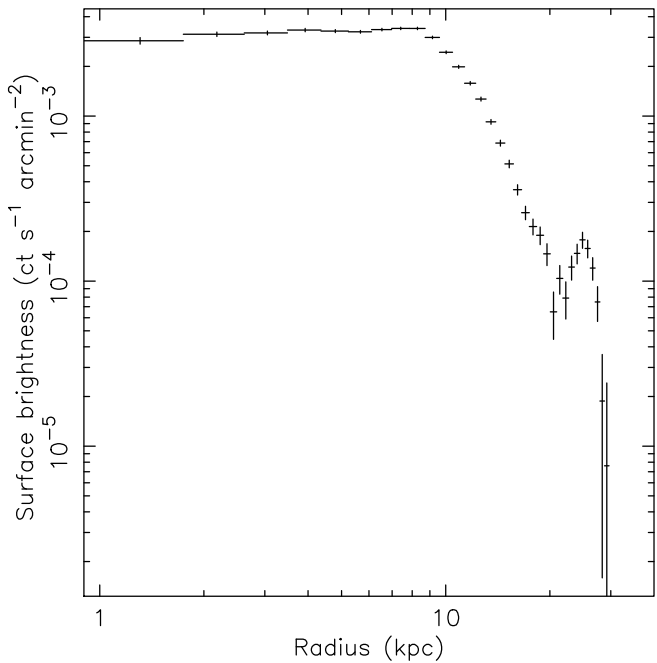

**Figure 9.** Surface brightness profile of the diffuse 0.1-1.9 keV flux (Fig. 5) about the centre of the system. The enhancement at $r \sim 25$ kpc is due to the X-ray extensions.

### 3.10 Overall properties and history

The *Antennae* has a total 0.1-2.0 keV luminosity (leaving the system) of $L_X = 1.1 \times 10^{41}$ erg s$^{-1}$, which is high compared to that of typical large spiral galaxies (Fabbiano 1989; David, Jones & Forman 1992; Read 1994). However a strong correlation is observed (Fabbiano, Gioia & Trinchieri 1988) between the X-ray and far-infrared luminosities among normal and starburst galaxies, and the X-ray luminosity of the *Antennae* is actually quite typical of high flux IRAS galaxies (Boller et al. 1992).

David et al. (1992) derive a relationship $L_X \propto L_{FIR}^{0.90-1.01}$ (90% confidence limits) from *Einstein* data, though this is based upon rough approximations for the X-ray spectra (each spectrum is assumed to be a 5 keV thermal plasma absorbed by a column of $3 \times 10^{20}$ cm$^{-2}$). This approximation may be appropriate for low activity spirals, which are dominated by emission from X-ray binaries, but in the case of the *Antennae*, which has a much softer spectrum, it leads to a serious overestimate in its luminosity – David et al. (1992) quote an (0.5-4.5 keV) X-ray luminosity of $6.3 \times 10^{41}$ erg s$^{-1}$, whilst our *ROSAT* results lead to a value of $L_X(0.5 - 4.5 \, \text{keV}) = 0.9 \times 10^{41}$ erg s$^{-1}$. The latter value compares well to the expected value from their X-ray/FIR correlation of $1.3 \times 10^{41}$ erg s$^{-1}$.

The conclusion that the *Antennae* actually has a *normal* X-ray luminosity for its activity, is confirmed by the fact that it falls close to the $L_X:L_{FIR}$ relationship derived for high FIR flux galaxies by Boller et al. (1992), and to that derived with *ROSAT* for a sample of nearby normal spiral galaxies by Read (1994).

What of the way in which this X-ray emission is distributed? The soft X-ray luminosity can be broken down into the following components: the discrete features A-G contribute 55%, the diffuse disc emission 41% (38% for the soft component attributed to hot gas, and 3% to the hard component from unresolved sources) and finally the two diffuse X-ray extensions account for 4% of the total. Separation of emission into discrete and diffuse components is sensitive to spatial resolution, and hence to source distance, and will also vary according to the analysis techniques employed. It appears from a variety of studies with *ROSAT* (Walterbos et al. 1993; Bregman & Pildis 1994; Pietsch et al. 1994; Read 1994; Heckman 1993; Vogler, Pietsch & Kahabka 1994; Schlegel 1994) that the fraction of genuinely diffuse emission (as distinct from harder emission from unresolved sources) varies quite widely in spiral galaxies, covering a range $\sim 10 - 80\%$, but that a fraction $\sim 50\%$ is common, particularly in galaxies experiencing high star formation rates. The temperature of such diffuse emission is typically confined to the rather narrow band $0.3 - 0.6$ keV. Hence on all these counts, the X-ray properties of the *Antennae* seem unremarkable. The only property which is unusual, is the morphology of the diffuse emission; combining a hot gas halo surrounding the galactic discs, with what appears to be a distorted but well-collimated outflow.

On the basis of the considerations above, one can construct a tentative picture for the evolution of the X-ray properties of the system as follows. It seems likely that the X-ray extensions correspond to a wind driven from one or both galaxies at the time, $\sim 1 - 2 \times 10^8$ years ago, when intense star formation was first triggered in their nuclei by the interaction. The bubble of very hot gas generated by the nuclear starburst will have broken out of the galactic disc perpendicular to the plane in the usual way, and once such a channel has been established, the wind should continue to flow so long as gas inflow within the disc continues to fuel further nuclear star formation. However the orientation of the outflows has been distorted from the normal oppositely directed bipolar form by the strong dynamical interaction of the two galaxies. This interaction has also been responsible for triggering distributed active star formation within the galactic discs, seen in the widespread H$\alpha$ knots and also in the presence of luminous X-ray sources within the discs arising from emission from superbubbles and massive X-ray binaries. This star forming activity generates localised outflows of hot gas into the galactic corona through 'chimneys' (Norman & Ikeuchi 1989), and this, together with further mixing and heating due to collisions between the gas associated with each galaxy, has generated the hot corona which envelopes the system.

## 4 CONCLUSIONS

The observations reported here represent the first X-ray spectral study of a major galaxy merger in progress with sufficient spatial resolution to resolve the various contributors to the emission. The *Antennae* has a high X-ray luminosity compared to normal spirals, but falls well on the $L_X$:$L_{FIR}$ relationship established for late-type galaxies. A number of discrete components to the X-ray emission have been identified, and some of these have counterparts at other wavelengths. X-ray sources are seen near the two galactic nuclei, but appear to derive from hot extranuclear gas rather than from the central starbursts themselves. In addition, X-ray emission is seen from two or three of the prominent H$\alpha$ knots within the galactic discs.

Nearly 50% of the flux associated with the system is diffuse, and its spectral properties suggests that the majority of this arises from hot gas rather than unresolved sources. The temperature of this gas is typically $4 \times 10^6$ K, and the total mass involved may be over $10^9 M_\odot$, depending upon its filling factor. The bulk of the diffuse emission covers the optical discs of the two galaxies, and appears to envelope the whole optical system, apart from the tidal tails. This halo is cooler and less extensive than those typically seen around bright ellipticals but is substantially brighter than halos seen in normal spiral galaxies. An enhancement in the surface brightness and temperature of the diffuse emission is seen in the region where the two discs are believed to be colliding. This may result from direct heating of gas by the collision, or from star formation triggered in the vicinity.

Two elongated structures extend to the NW and SW from the galactic discs to radii of $\sim 30$ kpc. It seems most likely that these indicate the presence of galactic winds, with a morphology which has been disturbed from the normal bipolar form by the rapid dynamical evolution of the system. They appear to contain approximately 10% of the total hot gas present, and could represent a significant loss of mass and energy from the system, though not of heavy elements if the low metallicities we infer are correct.

Galaxy mergers such as we are witnessing in the *Antennae* may have been much more common in the past (e.g. Larson (1990) and references therein) and could even be responsible for a large fraction of elliptical galaxies. In this context it is interesting that we see what may be a hot gas halo in the making, and also substantial mass loss in a wind. Similar observations of systems in more advanced merger stages are required to chart the evolution of the hot gas through the transformation from spiral to elliptical galaxies.

## ACKNOWLEDGEMENTS

AMR acknowledges the receipt of a SERC/PPARC studentship. We thank Mr. B.W.Hadley for producing the co-added optical image of the *Antennae* field, and the referee for helpful suggestions. Data reduction and analysis were performed on the Starlink node at Birmingham.